# On the Equation of State for Diquark Systems and their importance in Astrophysics and Cosmology

Santosh K. Karn[a]

New Delhi / PCCS, UP Tech. University, India
e-mail : skarn03@yahoo.com

**Abstract**
With a view to studying diquark formation in QCD phase transition and its consequences in high energy density physics, the energy per quark for the extended scalar diquarks (ESD) as a function of density is calculated, within the framework of an effective $\phi^4$ – theory, for several values of the effective interaction parameter ($\lambda$). Various equations of state (EOS) for the ESD systems corresponding to the different values of $\lambda$ are obtained. The importance of the EOS for such matter under extreme conditions of density and/or temperature is discussed and the role of the 'astrophysical diquarks' in astrophysics and 'cosmological diquarks' in cosmological conditions is also highlighted.

## 1. Introduction

According to Quantum Chromodynamics (QCD), the basic building blocks of nucleus can be described in terms of their constituents, namely quarks and gluons [1]. At high density/temperature of the order of 2 (GeV/fm$^3$)/200 MeV, the hadronic systems can undergo a transition from the hadronic to the deconfined state of quarks and gluons in which the quarks behave like free Fermi gas. The deconfined quarks and gluons in plasmonic state with respect to the colour charge is called quark.- gluon plasma (QGP) or QCD Plasma. Considerable progress has been made to the study of this state of matter and its properties in high energy density physics, particularly in the core of neutron stars. However, in spite of the developments over the years, it appears impossible [2] to predict with confidence the interior constitution of neutron stars.

It is to note here that the deconfined quarks and gluons may not be non-interacting rather, the quarks pair-up non-perturbatively mainly because of spin-spin interactions in order to lower the energy of the corresponding system. This may result in a creation of a composite system of two quarks called diquark (the word 'diquark' and its idea are due to Gell Mann [3] in 1964, and is a composite system of two quarks considered collectively and it has quantum numbers of a two quark system.) or of more number of quarks called multiquark system (proposed by Jaffe [4] about 30 years ago and several models namely, the MIT bag model [4], potential models [5], and flux tube models [6] etc.. have been put forward to study the multiquark systems); and the diquark has been studied by many others [5-20]. This kind of pairing of quarks, in general, allows the possibility of formation of other coloured-multiquark states which in conjunction with the available free gluons can give rise to a variety of QCD plasmas at several intermediate stages between the deconfined quarks-gluons phase and the hadronic phase. These intermediate phases namely, the diquark-quark-gluon (DQG) phase, diquark-gluon (DG) phase etc. may also exist in nature in an independent form or else may culminate into the so called colour-singlet hadronic systems. The possibility of formation of such (multiquark) states with large number of quarks has been shown by the bag model, the potential models support a few such states whereas the flux tube models suggest their instability. Out of these different multiquark states such as diquarks [3], tetraquarks [8], pentaquarks [9] and hexaquarks i.e., H-baryons [4], heptaquarks [10] etc., the study of diquarks has become of vital importance these days, particularly in the QCD phase transition. The possibility of the existence of diquarks in QCD was first pointed out and explored [11] by Ekelin and Fredriksson and later on emphasied by others [12-20]. The concept of diquarks in QGP has been introduced by Donoghue and Sateesh (DS) [14]. It has been argued by Kastor and Traschen [15] that the diquark phase is possible only if the transition from confined states of strongly interacting matter in hadrons to its deconfined state takes place. It is to emphasize that the existence of diquarks or their non existence ( at still higher temperature), and the imprints/signatures left by them would be very significant probes for the existence of QCD plasma. In this regard, some attempts have been made by Sateesh [14], H. Miao, et al. [21], H.M.Z. Ma,C. Gao [22] etc.. In addition to this, the recent discovery [23] of $\theta^+$ (1540) baryon a pentaquark, which may be considered as a two highly correlated ud-diquark pairs and an anti strange quark [9] has unfurled new excitement in this interesting field of research.

---

[a] Permanent address(personal): Flat # 327, Block # H 3, Sector- XVI, Rohini, New Delhi-110085, India.

Moreover, in the literature such quark-pairing is also considered like the formation of Cooper pairs in the superconductors, with diquarks being the 'Cooper pairs of QCD' [11,24]. The idea of superconducting quark matter is due to Barrois [25] and for the works in this context we refer to the review by Bailin and Love [24] and the work by Shuryak, Wilczek and collaborators [26]. In fact, the quarks are said [13] to have 'un-natural pairing up' to form diquarks and such pairing is possible[19] even in the presence of repulsive forces among the quarks if some of the forces there are more repulsive than the other. This suggests weak binding forces among the quarks but the quarks also have the colour degrees of freedom besides having a bound state (the diquarks).

In this context in the present work, we study in detail the extended scalar diquarks (ESD) systems within the framework of an effective $\phi^4$- theory. For this we briefly discuss the model we use here to study the diquark systems in Sect. 2. In this section, we calculate the energy per quark for the ESD as a function of density for several, positive and negative values of the effective interaction parameter ($\lambda$). In Sect. 3 we discuss the importance of the equations of state for the matter under extreme conditions of density and/or temperature and obtain the various equations of state for the ESD corresponding to the values of $\lambda$ considered in Sect. 2. We also highlight the role of diquarks in astro-physical situations and cosmological conditions in this section. Finally, the results are summarized and concluding remarks are made in Sect. 4.

## 2. The Diquark Model

Having briefly discussed the works on diquarks in the preceeding section we now discuss the diquark model we use here in the present work. As a matter of fact, the properties of a diquark is based on properties of the quarks of which the diquark is fromed. Quarks are colour triplet and have spin ½ and by accounting for spin, isospin, and colour degrees of freedom, we have four types of diquarks namely, ($\bar{3}$, 0), ($\bar{3}$, 1), (6, 0) and (6,1). The diquarks have SU (3) sextet for vector (spin one) diquarks and SU (3) triplet for scalar (spin zero) diquarks. It is known [11] that out of all possible states of diquarks, only the colour antitriplet scalar diquarks are energetically favoured. It is to mention that diquarks can be described [14] by an effective Lagrangian for a colour triplet field $\phi$,

$$L_{eff} = (1/2)(\partial_\mu \phi^+ \partial^\mu \phi - m_D^2 \phi^+ \phi) - \lambda(\phi^+ \phi)^2 \tag{1}$$

Here $m_D$ is the mass of the diquark and the effective parameter (self coupling constant) $\lambda$ is fixed [14] as 27.8 on the basis of P-matrix method of Jaffe and Low [27]. Karn; and Karn et al. [17-20] and several others [15,16] have used this for the study of diquark systems. The basic assumption of the model is that the bound diquarks in nucleon retain the same properties in a medium like QCD plasma. Karn; and Karn et al. have explored [17-20] the extended character of the diquark. They consider the Gaussian distribution function of the ESD as

$$F(k) = (N/2 \, (2\pi a^2 m_D^2)^{3/2}) \, ((k^2/b^2) + 1)^{-2} \cdot \exp(-k^2/2am_D^2) \tag{2}$$

and obtain the energy of the ESD gas as

$$E_D = \int d^3k (k^2 + m_D^2)^{1/2} F(k) + (\lambda/2V)[\int d^3k (k^2 + m_D^2)^{-1/2} F(k)]^2 \tag{3}$$

where N is the number of quarks, a is the Gaussian width parameter, b (= $2/B_{OC}$; with $B_{OC}$ like the first Bohr-radius) is 917 MeV. The energy of the diquarks can be expressed [17-20] in terms of quark number density $\rho$ ($\equiv N/V$) as

$$E_D = \frac{Nm_D}{4\pi}\left[2\sqrt{2\pi}\, a^{-3} I_1 + \lambda a^{-6}(\rho/m_D^2) I_2^2\right] \tag{4}$$

where

$$I_1 = \int_0^\infty dk' k'^2 (k'^2 + 1)^{1/2} \tilde{g}(k') \text{ with } \tilde{g}(k') = g(k')\exp(-k'^2/2a^2) \text{ and } g(k') = ((m_D k'/b)^2 + 1)^{-2}.$$

Here the variable k is replaced by ($m_D k'$) in eq. (4) for dimensional considerations and the integral $I_2$ is obtained by multiplying the integrand of $I_1$ by $(k'^2 +1)^{-1}$. For detail discussion on this we refer to our earlier works [17-20].

It is to emphasize here that the value of $\lambda$ is calculated by DS [14] is 27.8 and has been used by many others [15-20]. The energy of the diquark system is a direct function of $\lambda$ (c.f. eq. (4)) and it gives the quantitative picture of the diquark energy. In fact it is a measure of strength of repulsion that avoids Bose condensate at finite density. Further it is also said [16] that value of the $\lambda$ is greater by a factor of four than that obtained by the usual field theoretical value. It is for this reason we investigate in detail the effect of variation of $\lambda$ on the diquark energy. Here for computation of the diquark energy for various positive and negative, values of $\lambda$ the integrals $I_1$ and $I_2$ are computed numerically and the energy per quark for the ESD gas is calculated for various positive and negative values of the coupling constant $\lambda$ from eq. (4). The results are presented in Table 1 and shown in Fig. 1.

## 3. The Equation of State for the diquark systems and their importance

Having discussed the model in the previous section, we know proceed to discuss the importance of the equations of a state (EOS) for the diquark systems and obtain various EOS for the ESD systems. In this context, it is to mention that the study of the EOS for such a hot and highly compressed matter is interesting in itself and is very important these days in studying the astrophysical situations and cosmological conditions. In fact the study of such a hot and compressed matter, stars with that matter, stars with that matter in the core has emerged as excellent tools for studying the fundamental properties of gravity and matter under extreme conditions. The thermal emission from such compact stars may provide information about the history of evolution of such stars. In fact, we now feel the blooming of the exciting research field in which the physics, and the astrophysical and cosmological implications of such high density matter namely, the quark and/or diquark matter, play a fundamental role. However a part of this physics is based on the theories that are well tested in terrestial laboratories, a considerable part of it is still unknown [2]. It is to eimphasize that the formation of the diquarks during the QCD phase transition might have delayed the hydronization. It is in this regard the study of the EOS for the diquark systems and their importance has become of paramount importance. For this we here briefly discuss the model we use for studying the EOS for the diquark system. Karn;and Karn et al. use the relation, $P = - (\partial E_D/\partial V)_N$ and eq. (4), and express the pressure of the ESD gas as

$$P = [\vartheta_2 (3\vartheta_2 - 2\vartheta_4)(\lambda/2) + V\vartheta_3] / 3V^2 \qquad (5)$$

where

$$\vartheta_2 = \int d^3k \, (k^2 + m_D^2)^{-1/2} F(k), \quad \vartheta_3 = \int d^3k \, k^2 (k^2 + m_D^2)^{-1/2} F(k)$$

and $\vartheta_4 = \int d^3k \, k^2 (k^2 + m_D^2)^{-3/2} F(k)$.

The pressure of ESD gas in terms of the diquark matter density ($\rho_D \equiv m_D N/2V$; V is the volume occupied by the quarks) is

$$P = x_1 \vartheta'_2 \cdot \rho_D + 1.5 x_1 \lambda \, \vartheta'_1 (3\vartheta'_1 - 2\vartheta'_3) \cdot \rho_D^2 \qquad (6)$$

where $x_1 = (1/6\sqrt{2\pi}) a^{-3} m_D^{-4}$,

$$\vartheta'_1 = \int_0^\infty dk \, k^2 (k^2 + m_D^2)^{-1/2} \tilde{g}'(k); \; \tilde{g}'(k) = g'(k) \exp(-k^2/2m_D^2 a^2), g'(k) = ((k/b)^2 + 1)^{-2}$$

and the integrals $\vartheta'_2$ and $\vartheta'_3$ are obtained by multiplying the integrand of $\vartheta'_1$ by $k^2$ and

$k^2(k^2 + m_D^2)^{-1}$ respectively.

We calculate the pressure for ESD as a function of diquark matter density corresponding to the various positive and negative, values of the effective interaction parameter by λ using eq. (6). The integrals $\vartheta'_1$, $\vartheta'_2$ and $\vartheta'_3$ are computed numerically and the recent results thus obtained are presented in Table 2 and shown in Fig. 2. For details of the computational part we refer to our earlier works [17-20].

It is to mention here that the study of diquarks has become of considerable interest in high energy density physics, particularly in astrophysics and the study of primordial plasma at the Big Bang where diquarks might have delayed the hadronization. This novel matter is also expected to exist after the little bang under laboratory conditions, that is, through the present and planned relativistic heavy ion collisions (RHIC). It is to emphasize that in addition to this, the recent discovery of exotic pentaquarks namely, $\theta^+$ (1540) baryon which is considered [9] as two diquarks and an antistrange quark, has brought new excitement in the field. In fact, such a state of matter is believed to have existed for a fraction of a second after the big bang when our universe was created. This state of matter can as well be in the stars with such a matter or in the core of compact stars eg. neutron stars, hybrid stars etc.. There exists a possibility of a distinct family of quark and/or diquark stars having central densities higher than that of neutron stars.

In the context of 'astrophysical diquarks' in astrophysics it is to mention that the explanation for the formation of the compact stars or black holes by gravitational collapse of stars and the core collapse supernova explosions require the formation of such matter and it has become active areas of on going research these days. In fact, diquark has been suggested to play an important role in (i) a quark star, which may appear as dark matter [28], (ii)stars with diquark and/or quark matter with or without a hadronic envelop [15-20] e.g. Hybrid/neutron star;  (iii) supernova collapse or a 'hypernova' gamma ray burster, where diquarks may trigger neutrino bursts and the bounce-off; and (iii) the primordial plasma at the Big Bang, where diquarks might have delayed the hadronization.

We now highlight the possible roles of the diquarks in cosmological situations. It is to note here that the role of QGP in cosmology has been widely discussed in the literature [29(Schramm,Turner; Alcock et al., ICPAQGP proceedings)]. Karn; and Karn et al.[17-20] have found that energy of the ESD gas in a QGP is lower than that of the quark and / or point like diquark gas. This shows the stability of ESD matter. Consequently Karn [17] advocates the formation of ESD phase in a QGP, strongly argues that ESD phases have occurred during the early phases of the origin of the universe, and consequenty it plays an important role particularly during the first few fractions of a second, history (the Era) in the evolution of the universe. Thus the role of ESD phase in this context can be visualized and it is argued that the 'cosmological diquarks' might have played important roles in arriving at some meaningful consequences in cosmological studies [17-20] in the context of early universe. Fredrikksson et al. also have found that the QGP might suddenly enters a diquark phase where after just before the bouncing [30] a free-quark phase is again favoured. They argue that the transition into a diquark phase in a QGP results into neutrinos, as in the case of cosmic microwave background. The neutrino energy would have been of the order of the energy due to diquark correlation which is now cooled to KeV energies. They are now studying [31] the 'cosmological diquarks' in the primordial plasma at the Big Bang, where the diquarks might have delayed the hadronization. For a detailed discussions and references on the astrophysical and cosmological implications of the diquarks we refer to the work by Karn [19].

## 4. Summary and concluding remarks

With a view to understanding the various possible roles of diquark systems in QCD phase transition in high energy density physics particularly in astrophysical situations and cosmological conditions, in the present work, we have studied in detail the sensitivity of ESD energy on the effective interaction parameter ($\lambda$) within the framework of $\phi^4$ – theory. In this context, energy per quark for ESD system as a function of density has been calculated for various values of $\lambda$. Various equations of state for the ESD systems corresponding to the different values of $\lambda$ have been obtained. The importance of the EOS for such matter under extreme conditions of density and / or temperature is discussed and the role of the 'astrophysical diquarks' in astrophysics and 'cosmological diquarks' in cosmological conditions is also highlighted. It is to note that the study of the EOS of such matter is thus important in itself and is very important for studying the astrophysical situations as well as the dynamics of the big bang. Further, more theoretical study and experimental works are desirable to precisely study the role of the diquarks in the QCD phase transition in the context of the astrophysical conditions, particularly the compact stars, and the cosmological situations. It is to mention here that these equations of state can be solved by using TOV equations which will lead to different ESD stars. Thus, in this context different ESD stars and their properties can be explored in this framework. We hope to address some of them in future.

**Acknowledgement**

The author has been benefited by exchanging the views with the international stalwarts on this topic in the QFEXT05, Barcelona, Spain and in the international workshop on 'The New Physics of Compact Stars' at the European Centre for Theoretical studies in nuclear physics and related areas (ECT*), Trento, Italy. The author acknowledges the financial help from the organiser of the QFEXT05. He acknowledges the financial help for working as 'visiting scientist' at the ECT* and at the IEEC, UAB, Catalunya, Spain; and financial help for visiting the Service de Physique Theorique (SPhT), by the SPhT, Centre CEA de Saclay, 91191 Giff-Sur-Yvette Cedex, Saclay, France. He also acknowledges R.S. Kaushal and Y.K. Mathur for discussions on the topic; and the President and the Principal of PCCS for providing facilities.

**Table 1 : Energy per quark for the ESD systems for various values of $\lambda$**

| S. No. | Values of $\lambda$ | Energy per quark for the ESD systems for the following values of density $(\rho/m_q^3)$ | | | | |
|---|---|---|---|---|---|---|
| | | $(\rho/m_q^3) = 0.713 \times 10^{-2}$ | $(\rho/m_q^3) = 5.26 \times 10^{-2}$ | $(\rho/m_q^3) = 1.49 \times 10^{-2}$ | $(\rho/m_q^3) = 0.52 \times 10^{-2}$ | $(\rho/m_q^3) = 0.14 \times 10^{-2}$ |
| 1. | 27.80 | 0.719 | 0.688 | 0.624 | 0.613 | 0.4895 |
| 2. | 15.00 | 0.664 | 0.645 | 0.613 | 0.606 | 0.4885 |
| 3. | 11.38 | 0.648 | 0.635 | 0.610 | 0.605 | 0.4880 |
| 4. | 6.95 | 0.629 | 0.621 | 0.606 | 0.603 | 0.4880 |
| 5. | 6.18 | 0.626 | 0.619 | 0.605 | 0.602 | 0.4880 |
| 6. | 6.00 | 0.625 | 0.617 | 0.604 | 0.602 | 0.488 |
| 7. | 3.08 | 0.613 | 0.609 | 0.602 | 0.601 | 0.487 |
| 8. | 1.54 | 0.606 | 0.604 | 0.601 | 0.600 | 0.487 |
| 9. | – 15.00 | 0.535 | 0.551 | 0.586 | 0.592 | 0. 486 |

**Table 2 : New results for the equations of state for ESD systems**

| S. No. | ESD density ($\rho_D$) in g/cm³ | Pressure (P) for ESD system for different values of $\lambda$, in dyne/cm² | | | | | | | |
|---|---|---|---|---|---|---|---|---|---|
| | | $\lambda = 15.00$ | $\lambda = 11.38$ | $\lambda = 6.95$ | $\lambda = 6.18$ | $\lambda = 6.00$ | $\lambda = 3.08$ | $\lambda = 1.54$ | $\lambda = -15.00$ |
| 1. | $3.858 \times 10^{15}$ | $3.033 \times 10^{36}$ | $2.301 \times 10^{36}$ | $1.405 \times 10^{36}$ | $1.249 \times 10^{36}$ | $1.213 \times 10^{36}$ | $6.227 \times 10^{35}$ | $3.113 \times 10^{35}$ | $-3.033 \times 10^{36}$ |
| 2. | $2.787 \times 10^{15}$ | $1.582 \times 10^{36}$ | $1.200 \times 0^{36}$ | $7.329 \times 10^{35}$ | $6.517 \times 10^{35}$ | $6.327 \times 10^{35}$ | $3.248 \times 10^{35}$ | $1.624 \times 10^{35}$ | $-1.582 \times 10^{36}$ |
| 3. | $1.715 \times 10^{15}$ | $5.990 \times 10^{35}$ | $4.545 \times 10^{35}$ | $2.776 \times 10^{35}$ | $2.468 \times 10^{35}$ | $2.396 \times 10^{35}$ | $1.230 \times 10^{35}$ | $6.15 \times 10^{34}$ | $-5.990 \times 10^{35}$ |
| 4. | $1.393 \times 10^{15}$ | $3.955 \times 10^{35}$ | $3.000 \times 10^{35}$ | $1.832 \times 10^{35}$ | $1.629 \times 10^{35}$ | $1.582 \times 10^{35}$ | $8.120 \times 10^{34}$ | $4.06 \times 10^{34}$ | $-3.954 \times 10^{35}$ |
| 5. | $1.072 \times 10^{15}$ | $2.340 \times 10^{35}$ | $1.775 \times 10^{35}$ | $1.084 \times 10^{35}$ | $9.640 \times 10^{34}$ | $9.360 \times 10^{34}$ | $5.805 \times 10^{34}$ | $2.402 \times 10^{34}$ | $-2.340 \times 10^{35}$ |
| 6. | $6.859 \times 10^{14}$ | $9.584 \times 10^{34}$ | $7.271 \times 10^{34}$ | $4.441 \times 10^{34}$ | $3.949 \times 10^{34}$ | $3.834 \times 10^{34}$ | $1.968 \times 10^{34}$ | $9.84 \times 10^{33}$ | $-9.584 \times 10^{34}$ |
| 7. | $5.359 \times 10^{14}$ | $2.34 \times 10^{34}$ | $4.438 \times 10^{34}$ | $2.710 \times 10^{34}$ | $2.410 \times 10^{34}$ | $2.340 \times 10^{34}$ | $1.201 \times 10^{34}$ | $6.006 \times 10^{33}$ | $-5.850 \times 10^{34}$ |
| 8. | $5.066 \times 10^{14}$ | | $3.967 \times 10^{34}$ | $2.423 \times 10^{34}$ | $2.154 \times 10^{34}$ | $2.091 \times 10^{34}$ | $1.074 \times 10^{34}$ | $5.368 \times 10^{33}$ | $-5.228 \times 10^{34}$ |
| 9. | $3.483 \times 10^{14}$ | $2.471 \times 10^{34}$ | $1.875 \times 10^{34}$ | $1.145 \times 10^{34}$ | $1.018 \times 10^{34}$ | $9.885 \times 10^{33}$ | $5.075 \times 10^{33}$ | $2.538 \times 10^{33}$ | $-2.471 \times 10^{34}$ |
| 10. | $2.573 \times 10^{14}$ | $1.348 \times 10^{34}$ | $1.023 \times 10^{34}$ | $6.247 \times 10^{33}$ | $5.555 \times 10^{33}$ | $5.393 \times 10^{33}$ | $2.769 \times 10^{33}$ | $1.385 \times 10^{33}$ | $-1.348 \times 10^{34}$ |
| 11. | $1.429 \times 10^{14}$ | $4.161 \times 10^{33}$ | $3.157 \times 10^{33}$ | $1.928 \times 10^{33}$ | $1.715 \times 10^{33}$ | $1.665 \times 10^{33}$ | $8.546 \times 10^{32}$ | $4.274 \times 10^{32}$ | $-4.161 \times 10^{33}$ |
| 12. | $1.118 \times 10^{14}$ | $2.548 \times 10^{33}$ | $1.933 \times 10^{33}$ | $1.181 \times 10^{33}$ | $1.050 \times 10^{33}$ | $1.019 \times 10^{33}$ | $5.233 \times 10^{32}$ | $2.617 \times 10^{32}$ | $-2.548 \times 10^{33}$ |
| 13. | $9.191 \times 10^{13}$ | $1.721 \times 10^{33}$ | $1.306 \times 10^{33}$ | $7.975 \times 10^{32}$ | $7.091 \times 10^{32}$ | $6.885 \times 10^{32}$ | $3.535 \times 10^{32}$ | $1.768 \times 10^{32}$ | $-1.721 \times 10^{33}$ |
| 14. | $8.036 \times 10^{13}$ | $1.316 \times 10^{33}$ | $9.982 \times 10^{32}$ | $6.097 \times 10^{32}$ | $5.421 \times 10^{32}$ | $5.263 \times 10^{32}$ | $2.703 \times 10^{32}$ | $1.352 \times 10^{32}$ | $-1.315 \times 10^{33}$ |
| 15. | $5.043 \times 10^{13}$ | $5.181 \times 10^{32}$ | $3.931 \times 10^{32}$ | $2.401 \times 10^{32}$ | $2.135 \times 10^{32}$ | $2.073 \times 10^{32}$ | $1.065 \times 10^{32}$ | $5.328 \times 10^{31}$ | $-5.180 \times 10^{32}$ |
| 16. | $3.676 \times 10^{13}$ | $2.754 \times 10^{32}$ | $2.089 \times 10^{32}$ | $1.276 \times 10^{32}$ | $1.135 \times 10^{32}$ | $1.102 \times 10^{32}$ | $5.660 \times 10^{31}$ | $2.833 \times 10^{31}$ | $-2.753 \times 10^{32}$ |
| 17. | $2.926 \times 10^{13}$ | $1.744 \times 10^{32}$ | $1.323 \times 10^{32}$ | $8.084 \times 10^{31}$ | $7.189 \times 10^{31}$ | $6.980 \times 10^{31}$ | $3.586 \times 10^{31}$ | $1.796 \times 10^{31}$ | $-1.743 \times 10^{32}$ |
| 18. | $2.425 \times 10^{13}$ | $1.199 \times 10^{32}$ | $9.096 \times 10^{31}$ | $5.557 \times 10^{31}$ | $4.942 \times 10^{31}$ | $4.798 \times 10^{31}$ | $2.465 \times 10^{31}$ | $1.253 \times 10^{31}$ | $-1.198 \times 10^{32}$ |
| 19. | $1.338 \times 10^{13}$ | $3.648 \times 10^{31}$ | $2.768 \times 10^{31}$ | $1.692 \times 10^{31}$ | $1.505 \times 10^{31}$ | $1.461 \times 10^{31}$ | $7.513 \times 10^{30}$ | $3.770 \times 10^{30}$ | $-3.643 \times 10^{31}$ |

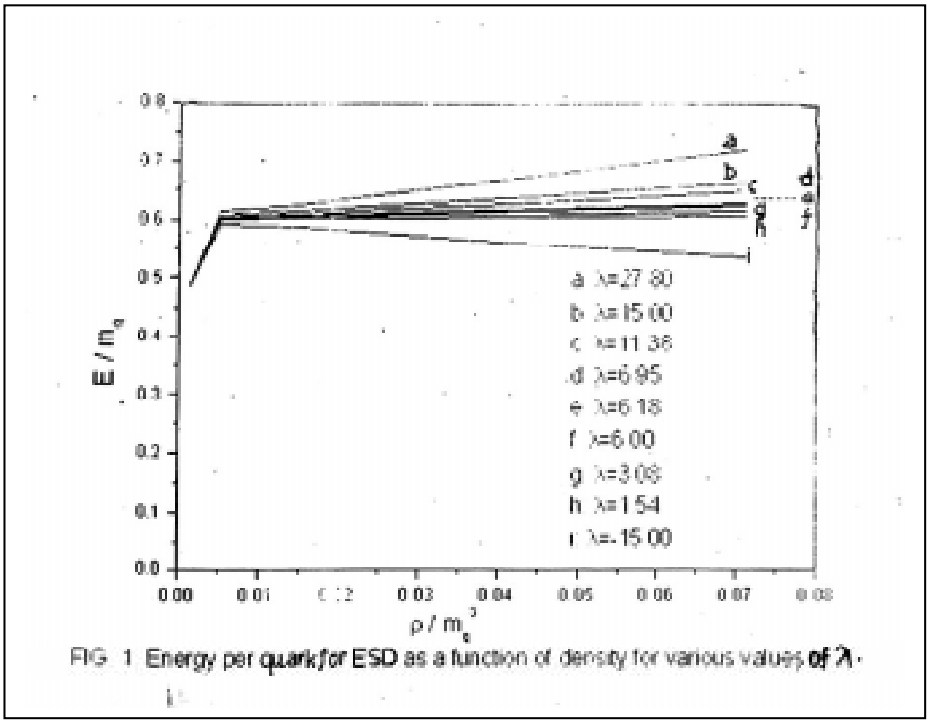

FIG. 1 Energy per quark for ESD as a function of density for various values of λ.

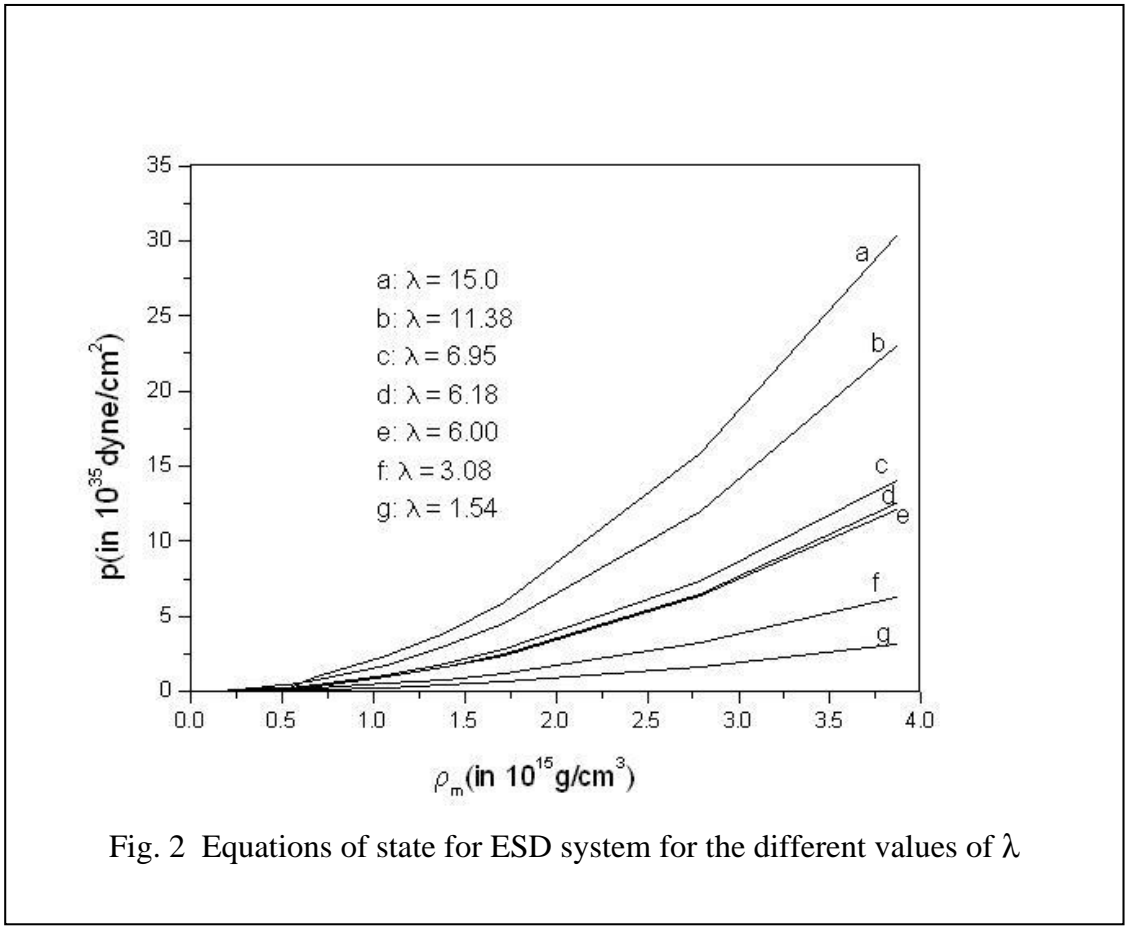

Fig. 2 Equations of state for ESD system for the different values of λ